# It Takes a Village: Multidisciplinarity and Collaboration for the Development of Embodied Conversational Agents

Danai Korre

School of Engineering, The University of Edinburgh, d.korre@ed.ac.uk

Embodied conversational agent (ECA) development is a time-consuming and costly process that calls for knowledge in a plethora of different and not necessarily adjacent disciplines. Engaging in activities outside of one's core research to acquire peripheral skills can impede innovation and potentially restrict the outcomes within the boundaries of those acquired skills. A proposal to tackle this challenge is creating collaborative communities of experts from the contributing disciplines to the field of ECAs that via clearly defined roles, expectations and communication channels can help extend the field of ECA research.

## 1 INTRODUCTION

The field of embodied conversational agents (ECAs) is multidisciplinary and constitutes a subcategory of conversational agents and conversational user interfaces.

The term "Embodied Conversational Agent" was coined by Justine Cassell in 2000 and is defined as follows:

> "[ECAs are] computer interfaces that can hold up their end of the conversation, interfaces that realize conversational behaviors as a function of the demands of dialogue and as a function of emotion, personality, and social conversation" [8]

In layman's terms ECAs are virtual characters with the ability to converse with a human through verbal (speech) and/or non-verbal communication (text and/or gestures) [9].

Previous research has identified the possibilities of using ECAs for tourism and culture [3,14], business applications [11,17,18], journalism [5], healthcare [12,27,28], as companions [6], as sales agents [1,7,13], for military training [19,23], for psychological support [20], for education [2, 10] and in various other roles.

Even though recent technological advancements are making the development of ECAs easier and relatively cost effective for non-experts, it is still a complex and time-consuming process that demands expertise and a wide array of skills. Additionally, the results are a long way from industrial level agents used in commercial games. It also limits research on the capabilities of the researcher as they spend time on tasks beyond their research focus which hinders their progress.

Previous research has also shown that individual ECA attributes can affect the interaction as information is conveyed by verbal, non-verbal and extra-linguistic channels [16]. A more advanced ECA (for example ECAs using multimodal input such as face recognition and natural language) can be more believable than their simplistic counterparts; the complexity of those agents though, comes with challenges as these systems are prone to mistakes (e.g., misinterpreting semantics of natural language) and demand more development time and expertise. One way to tackle these problems is to use more simplified approaches (e.g., decision tree mechanisms or simplistic graphics) but they also make for a less realistic experience. Trade-offs, such as the ones mentioned above, make finding the optimal approach in a specific setting a nontrivial task [16,21].

## 2  MULTIDISCIPLINARY PERSPECTIVES

Embodied conversational agents are the result of many contributing disciplines. Those disciplines differ for each ECA depending on its capabilities, purposes, and modes of interaction. For example, an ECA that uses speech input needs speech recognition and speech-to-text technology, while ECAs with text input do not. In general, ECAs are by their nature multidisciplinary as shown in Figure 1.

Each discipline contributes to specific ECA aspects. Computer science contributes to the aspects of natural language processing, artificial intelligence, image recognition and/or speech synthesis and recognition [4]. Linguistics contributes to dialogue design, speech patterns, semantics, dictionary and more [4]. Art and design contribute to the areas of 3D and/or 2D design, computer graphics, animation, character design, art, and overall visual representation [16]. Cognitive science, psychology, anthropology, and sociology contribute to motivation, perception, engagement, satisfaction, affordances, biases and more [4,24]. Finally, communication studies and interaction design contribute to user experience design, user interface design, usability, and accessibility [4].

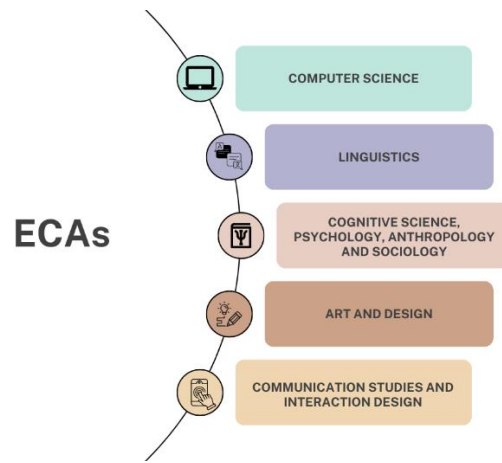

Figure 1 Contributing disciplines to the field of ECAs

## 3  COLLABORATION IN ECA DEVELOPMENT

Collaboration between researchers and practitioners from different disciplines is especially important in an academic setting where resources and time are often limited.

Interdisciplinary collaboration can lead to the extension of the ECA development boundaries by combining expertise from different fields to address domain-specific challenges. Due to their versatility, ECAs can be used in multiple domains and having multiple points of view can lead to innovation fostering, enhanced creativity as well as more natural interaction and increased usability.

Effective collaboration in ECA development projects can be challenging mainly due to resource limitations, lack of trust, communication hurdles or lack of common research interests and vision. However, it should be treated as an opportunity for multidisciplinary research, networking, and future collaborations. One way to overcome the challenges is by adopting strategies for effective collaboration such as trust building, ensuring accountability and productivity, clearly defined goals, and vision, providing regular feedback and effective communication channels [25]. These solutions can be applied to in person but also in remote collaborations that became even more prevalent in the post-pandemic era.

## 4 CASE STUDIES AND EXAMPLES

A real-world example of a successful collaborative project between an academic and an industrial partner in ECA development is Susa. Susa is a conversational agent that promotes teamwork and collaborative practices. The study provides empirical data on co-designing with end-users, based on the principles of design thinking. This project is a collaboration between the National Institute of Public Health from the University of Southern Denmark and Gnist Denmark, a company that translates behavioral design into practical solutions [26]. Regarding the authors' contributions, the statement demonstrates that the study design and project idea were developed collaboratively by the academic and industrial partners. The industrial partners were responsible for the development of the ECA and organized the user workshops. Data collection was a cooperation of members from both the academic and industrial partners, while the analysis was primarily conducted by the academic partners.

## 5 FUTURE DIRECTIONS AND IMPLICATIONS

The development of ECAs requires multidisciplinary collaboration in response to emerging technological developments. The increasing use of immersive technologies, such as virtual and augmented reality, presents new opportunities and challenges for ECA development. Research conducted on ECAs, whether in immersive applications [22] or more conventional media [15], highlights certain constraints that can be effectively mitigated through multidisciplinary teams relying on cross-school collaborations or collaborations between academia and industry. These limitations include among other things the animation or the visual representation of the agents which can affect the interaction [15,22].

Furthermore, alongside the rapid technological advancements, the emergence of ethical concerns and the need for inclusivity across diverse user populations add further complexity to the development lifecycle of ECAs.

Therefore, multidisciplinary collaboration can have an impact on ECA research as the individual development tasks can be distributed to parties with the relevant expertise while the researcher can focus on the task at hand.

## 6 CONCLUSION

In conclusion, multidisciplinary collaboration for the development of ECAs can alleviate the pressure of development by an individual. Beyond the need for expertise on multiple disciplines or acquiring peripheral skills to the main research focus, researchers need to respond to emerging technological developments as well. Previous examples have proved that multidisciplinary collaboration can be very beneficial and initiatives for promoting these collaborations should be considered in the ECA community.